\documentclass[11pt]{article}
\usepackage[backend=bibtex,style=chem-acs,sorting=none]{biblatex}
\usepackage[version=3]{mhchem} 
\usepackage[table,xcdraw]{xcolor}
\usepackage{amsmath, amsthm, amssymb, amstext, bm, mathtools}
\usepackage{siunitx}
\usepackage{authblk}
\usepackage{setspace}
\usepackage{braket}
\usepackage[margin=1in]{geometry}
\usepackage{float}
\usepackage{tabularx}
\usepackage{graphicx}
\usepackage{times}      
\usepackage{soul}
\usepackage{color}     
\usepackage{subcaption}
\usepackage{changepage}
\usepackage{dirtytalk}
\usepackage{array}
\usepackage{textcomp}
\usepackage{hyperref}
\hypersetup{
    colorlinks,
    citecolor=black,
    filecolor=black,
    linkcolor=black,
    urlcolor=blue
}
\usepackage{cleveref}
\usepackage{xr}
\usepackage{xparse}
\DeclareDocumentCommand\vectorbold{ s m }{\IfBooleanTF{#1}{\boldsymbol{#2}}{\mathbf{#2}}}
\DeclareDocumentCommand\vb{}{\vectorbold}
\newcommand{\tioo}{\text{TiO${}_2$}}
\usepackage{listings}
\usepackage[scaled]{helvet}  
\newcommand{\helvet}[1]{{\sffamily #1}}

\title{\textit{Ab Initio} Mechanisms and Design Principles for Photodesorption from TiO${}_{\vb{2}}$}

\author[1]{Aaron R. Altman}
\author[1,2,3]{Felipe H. da Jornada\thanks{\texttt{jornada@stanford.edu}}}
\affil[1]{Department of Materials Science and Engineering, Stanford University, Stanford, CA 94305, USA}
\affil[2]{Stanford Institute for Materials and Energy Sciences, SLAC National Accelerator Laboratory, Menlo Park, CA 94025, USA}
\affil[3]{Stanford PULSE Institute, SLAC National Accelerator Laboratory, Menlo Park, CA 94025, USA}

\begin{document}
\maketitle

\begin{abstract}
Photocatalytic reactions often exhibit fast kinetics and high product selectivity, qualities which are desirable but difficult to achieve simultaneously in thermally driven processes.
However, photo-driven mechanisms are poorly understood owing to the difficulty in realistically modeling catalysts in optically excited states.
Here we apply many-body perturbation theory (MBPT) calculations to gain insight into these mechanisms by studying a prototypical photocatalytic reaction, proton desorption from a rutile \tioo\ (110) surface. 
Our calculations reveal a qualitatively different desorption process upon photoexcitation, with an over 50\% reduction in the desorption energy and the emergence of an energy barrier.
We rationalize these findings with a generalizable model based on Fano theory and explain the surprising increase of excitonic effects as the proton detaches from the surface. 
Our model also yields a connection between how the alignment of relevant ionization potentials affects the shape of the excited-state potential energy surface.
These results cannot be qualitatively captured by typical constrained density-functional theory and highlight how contemporary first-principles MBPT calculations can be applied to design photocatalytic reactions.
\end{abstract}

\begin{refsection}
Photocatalysis encompasses a class of reactions in which photon energy directly drives or improves a catalytic reaction rather than standard thermocatalysis that is driven by heat and/or pressure. Microscopically, light excites a photocatalyst or molecule to an excited quantum state, altering the effective potential energy surface (PES) that the ions in the system experience, and allowing for qualitatively different energy barriers, ionic dynamics, reaction rates, and reactant selectivity. The potential applications of photocatalysis are diverse, ranging from new ways to synthesize ammonia to plastics decomposition to drug synthesis, and many others~\cite{han2021progress, chen2018photocatalytic, ouyang2021recent, trowbridge2018multicomponent, ameta2016photocatalysis}.
Recent experiments have demonstrated the dramatic effects of optical excitations in various systems. 
For instance, nanoparticle catalysts can display dramatically faster reaction rates and higher product selectivity among multiple products when illuminated relative to a thermally driven baseline~\cite{qi2020dynamic, zhou2018quantifying, shetty2020catalytic}. These reactions can also often be performed at much milder experimental conditions, in terms of temperature and pressure requirements, than their thermal counterparts~\cite{li2022nanostructured, hosseini2020visible}.
Additionally, studies have demonstrated the capability of photodriven reactions to access nonthermal products such as in the oxidation of CO to CO$_2$ on a ruthenium surface~\cite{bonn1999phonon, funk2000desorption}. 

There are numerous reports of different photocatalysts, reactions, and proposed mechanisms to explain the advantage of light-driven chemistry~\cite{zhang2019plasmon, daza2016co, qi2020dynamic, shetty2020catalytic, pan2020photons, yamazaki2015photocatalytic}. 
Two important classes of photocatalysts are plasmonic nanoparticles~\cite{Mukherjee2013, zhang2019plasmon, mascaretti2020hot, chen2023origin, seemala2019plasmon, boerigter2016mechanism, aslam2018catalytic, hartland2017s, zhou2018quantifying} and semiconductor surfaces, with \tioo\ an especially common choice for multiple reactions~\cite{duncan2007theoretical, Tritsaris2014dynamics, Berardo2014modeling, habisreutinger2013photocatalytic, agrios2005state}. 
Plasmonic nanoparticles host a variety of photocatalytic mechanisms, including direct plasmon-driven reactions, hot electron generation, photothermal effects, and others~\cite{carlin2023nanoscale}. Due to their strong light absorption and high surface area, nanoparticles are an ideal platform for realizing such photocatalytic processes in operational settings. However, their non-static geometry, metallic character, and strong size dependence often obscure the underlying photophysical processes, making it difficult to untangle competing effects such as non-thermal phonons and electrons, and direct energy transfer routes. On the other hand, while semiconductor photocatalysts do not absorb light as strongly as metallic nanoparticles at the plasmon resonance, they have the advantage of hosting much longer-lived excitations that are energetically discrete. Hence, one can often approximate the ionic dynamics in a photoexcited semiconductor catalyst as an adiabatic evolution given the lowest-energy excited-state PES, which is not possible in a metallic system due to the continuum of optically addressable states.

Even with the large experimental effort on both classes of photocatalysts, there is no general predictive theoretical understanding of the underlying mechanisms due to the difficulty in computing optically excited materials including the necessary electron-hole correlations and the structural complexity. 
While density-functional theory (DFT) is often used for catalytic studies, DFT is a ground-state formalism and cannot rigorously yield arbitrary excited-state properties of materials. Other methods based on DFT, such as the delta-self-consistent-field method~\cite{hellman2004potential}, time-dependent DFT~\cite{jin2022vibrationally, jin2023excited, butchosa2014carbon, wang2024excitonic}, and variants of molecular dynamics~\cite{melani2024effects, pham2017modelling, akimov2015makes, stier2002nonadiabatic, muuronen2017mechanism, tully1990molecular, akimov2013pyxaid, liu2020excitonic, jiang2021real}, can capture some of the important ionic dynamics in an electronic or optically excited system; however, typical approximations of locality and adiabaticity made to the exchange-correlation functional restrict the applicability of these methods. Moreover, while range-separated hybrid functionals are expected to capture some critical many-electron correlation effects~\cite{jacquemin2008td, wong2010optoelectronic, korzdorfer2014organic}, they still have to be systematically demonstrated in spatially inhomogeneous systems.
Another class of methods known as quantum embedding partition the system into a chemically active portion and an environment, treating the active region with an exact diagonalization approach, such as configuration interaction (CI), and the environment at a lower theory level, such as DFT~\cite{huang2011quantum, libisch2014embedded, Mukherjee2013}. 
While such methods can accurately capture the excited states in the active region, they have difficulties treating large-scale correlations due to the poor computational scaling of the exact diagonalization methods. 

Recently, first-principles many-body perturbation theory (MBPT) calculations have started to be applied to the study of photocatalytic systems~\cite{jin2020critical, zhang2022interfacial, zhou2021new, jin2018behavior, chen2018mediating, hu2018photocatalytic, biswas2021excitonic, fan2022type}. 
MBPT is a uniquely suited approach for the study of photocatalytic reactions as it rigorously treats electronic correlations and optical excitations, and recent algorithmic advances~\cite{altman2024mixed} and massively parallel codes~\cite{DelBen2020accelerating} suitable for large supercomputing clusters now admit explicit calculations on relevant photocatalytic systems with tens to hundreds of atoms. Earlier works based on MBPT calculations have revealed important insights into photocatalytic mechanisms of specific reactions, often related to band alignments or charge-transfer exciton character~\cite{jin2020critical, zhang2022interfacial, fan2022type}. Still, due to the high computational cost, such calculations were restricted to a few structural configurations and could not access the continuous dependence of the electron-hole correlations on the ionic coordinates. As we show, such effects are general, play a critical role in setting the efficiency of a photocatalytic reaction, and may provide unique experimental fingerprints for validating a photocatalytic mechanism.

In this work, we apply MBPT to a simple yet realistic photocatalytic reaction, the proton desorption from a rutile \tioo{} (110) surface, to understand in detail the underlying photophysics and produce generalizable principles for the rational design of other simple photocatalytic reactions. Notably, we demonstrate that the photocatalytic reaction is clearly understood by tracking the evolution of both the quasiparticle excitation energies and exciton wavefunctions for the whole excitation spectrum, quantities directly accessible with MBPT.

The photo-desorption of a proton from \tioo\ has several advantages as the object of a theoretical study. First, unlike more complex reactions that may have numerous reaction paths and competing factors in their mechanisms, the photo-desorption of a proton is well-approximated by a single, one-dimensional (1D) reaction coordinate given the energetics and symmetry of the lowest-energy adsorption site~\cite{kunat2004adsorption}. 
In particular, a 1D reaction coordinate circumvents the need, for now, to deal with another complex problem that is orthogonal to the calculation of excited-state PESs: the necessity of computing excited-state forces. 
Second, a 1D reaction pathway is easy to finely sample, from which one could observe changes such as the softening of phonon/vibrational modes and allows us to understand the adiabatic evolution of the excited-state wavefunctions, which is critical near the crossing of two PESs. 
Finally, the discrete excitation spectrum of \tioo\ admits the straightforward interpretation of the reaction evolution as adiabatic along the lowest excited-state PES without an intersystem crossing. 
To enable a thorough and self-contained understanding of this elementary reaction step, we do not consider the effects of defects such as the common oxygen vacancy that has been extensively documented to improve the photocatalytic performance of \tioo, which we comment on briefly in the Discussion section~\cite{nakamura2000role, li2017synergistic, bi2021tuning, zhao2015effect, chen2018facile, feng2018significantly, qi2018photocatalytic}.

In the following sections, we describe our computational approach, present the results of our calculations and the underlying physical picture, and introduce a model that allows us to obtain quantitative approximations to the \textit{ab initio} results. We also discuss how our model can be generalized to other simple photocatalytic reactions and demonstrate how it can form the basis of a rational approach to designing semiconductor photocatalysts.

\section{Results}
\subsection{Surface Structure}
\begin{figure}
    \centering
    \includegraphics[scale=0.5]{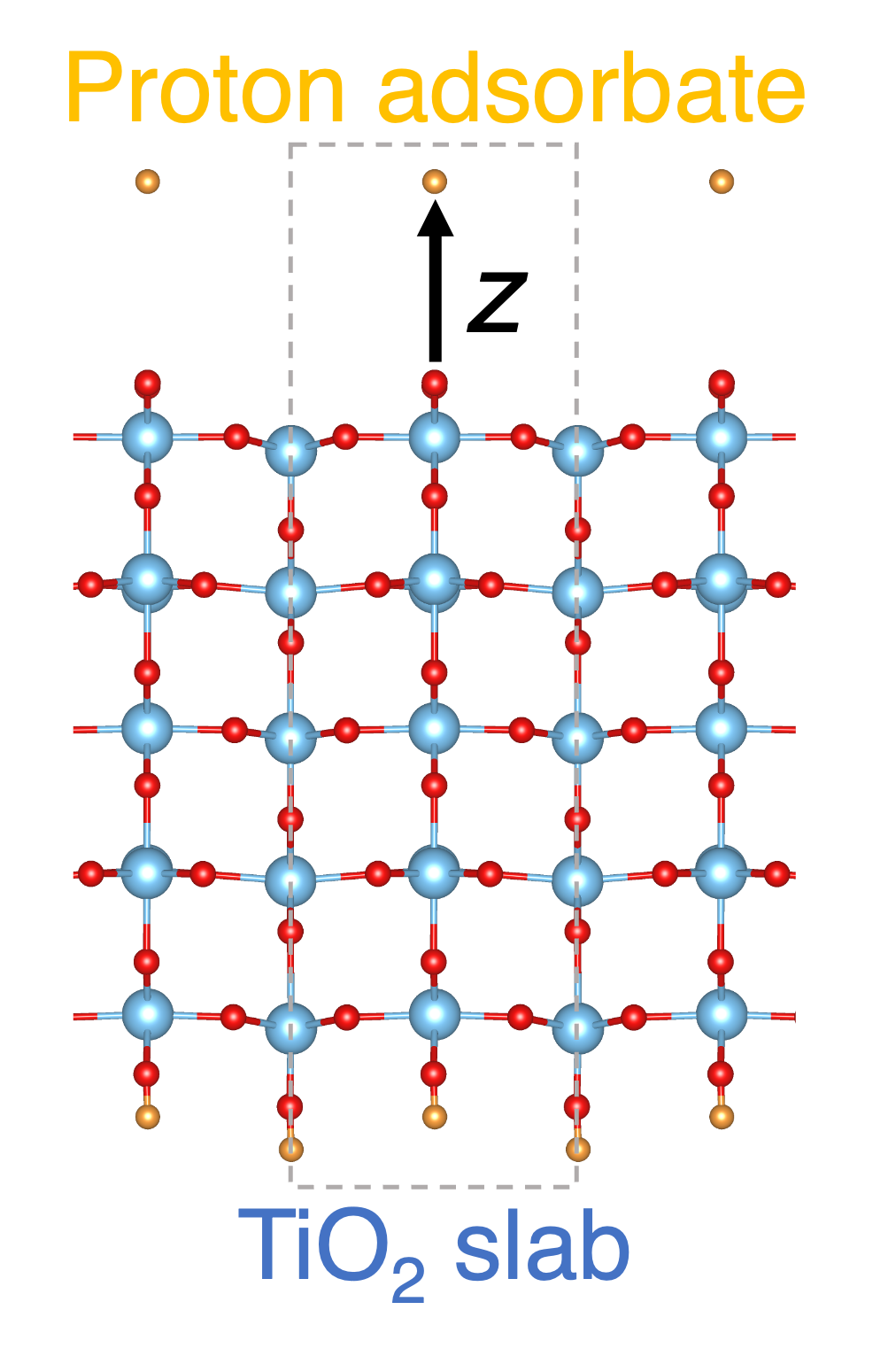}
    \caption{Relaxed structure of \tioo\ slab where red atoms are oxygen, blue atoms are titanium, and yellow atoms are hydrogen, where the ones at the bottom are fixed as a passivating layer. The adsorbed proton is shown at the top of the slab, and corresponds to 1 adsorbate per $2\times 1$ supercell. The arrow labeled $z$ denotes the reaction coordinate, \textit{i.e.} the distance between the adsorbed hydrogen and 2-fold coordinated oxygen O$_{2c}$. The thin dashed box is the unit cell (the vacuum of $\sim$25~Å is not shown).}
    \label{fig:struct}
\end{figure}

As in other studies, we simulate the reaction here by calculating the ground and excited-state energies of a series of structures along the reaction coordinate; for the desorption reaction, this entails moving the proton farther from the \tioo\ (110) surface. 
The choice of a proton adsorbate was partially motivated by the fact that this allows our system to always be in a closed-shell configuration.
We calculate the ground-state PES with DFT as the change in the total energy of each ionic configuration. The excited-state PESs are obtained by adding the excitation energies from MBPT to the ground-state energy at each configuration. 
The \tioo\ surface is constructed as a 5-layer, $2\times 1$ in-plane supercell slab (Fig.~\ref{fig:struct}), with hydrogen atoms added to passivate the dangling bonds on the bottom of the slab. The supercell was chosen to avoid interactions between adjacent adsorbates (see Supporting Information); treatment of the high-coverage limit requires separate studies.
Previous work has shown that the preferred adsorption site of a hydrogen atom is on the 2-fold coordinated oxygen atom O$_{2c}$ at the top of the slab~\cite{kunat2004adsorption}, so we use this as the equilibrium structure and simulate desorption by pulling the proton along the normal to the surface. 
The equilibrium structure is relaxed with the passivating hydrogens fixed to maintain the bulk structure. Given its small mass, the proton is expected to evolve much faster than the ions on the slab; accordingly, we do not relax the atomic configurations after we obtain the equilibrium ground-state configuration. 
Despite this, we also performed a separate set of calculations with this additional relaxation and found no significant differences in our computed PESs (see Supporting Information). 

We perform MBPT calculations within the GW and GW plus Bethe-Salpeter equation (GW-BSE) formalism to capture quasiparticle and optical excitations~\cite{hybertsen1986electron, rohlfing2000electron}. We compute the dielectric matrix and quasiparticle self-energy at the level of the random phase approximation and GW approximation. This captures image-charge and surface screening effects which are critical to obtain accurate single-particle energies and relative alignments that determine the qualitative behavior of the photocatalytic reaction. To obtain both neutral optical excitation energies and their wavefunctions, we solve the BSE. See the Methods below and the Supporting Information for all structures and convergence details for all calculations.

\subsection{\textit{Ab Initio} Calculations}
\begin{figure}[H]
    \centering
    \includegraphics[width=0.8\textwidth]{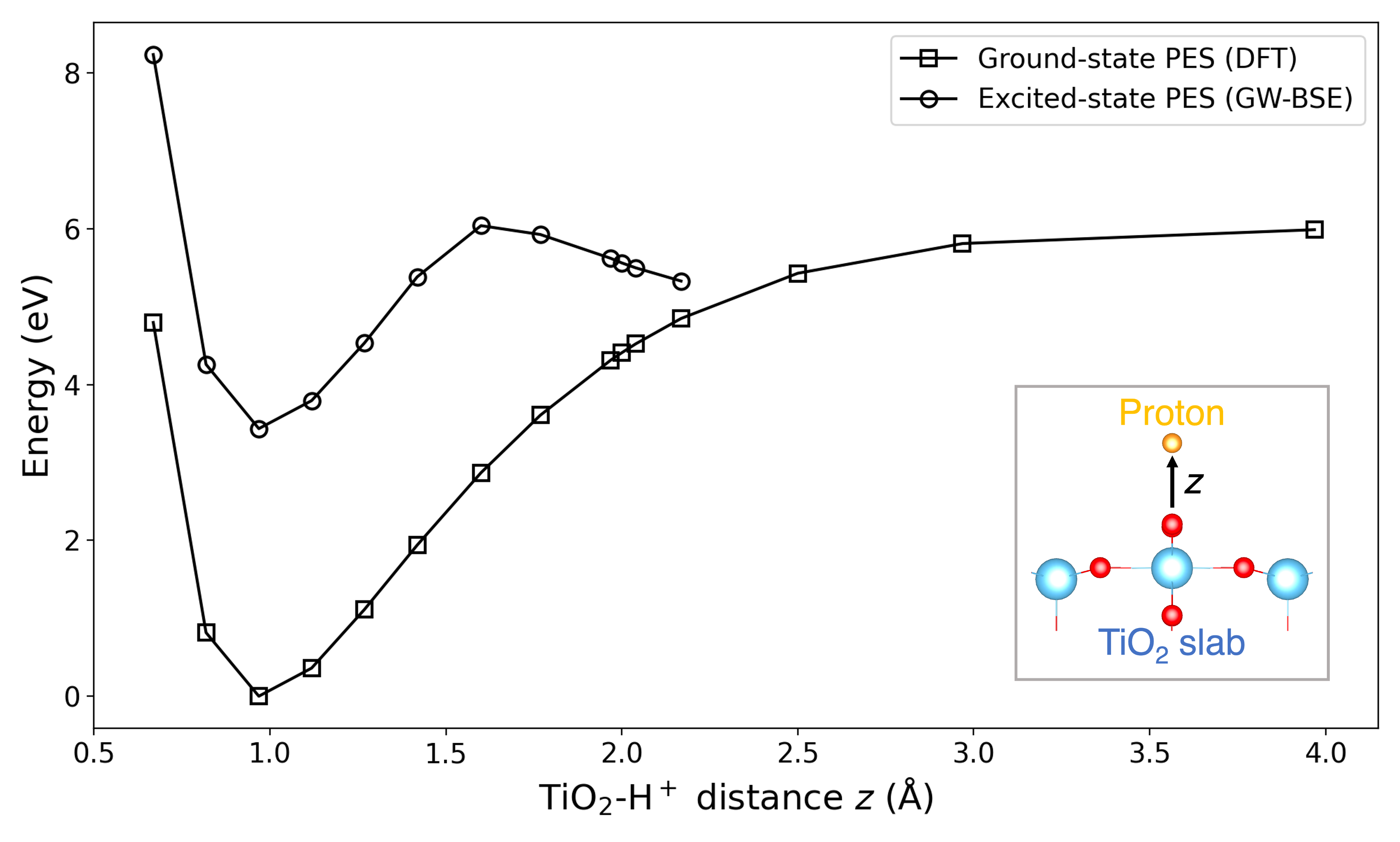}
    \caption{\textit{Ab initio} ground-state and lowest-energy excited-state PESs. Inset shows the reaction coordinate $z$. The excited-state PES is sampled right before the system acquires an open-shell, spin-polarized character (see Supporting Information)}.
    \label{fig:XPES}
\end{figure}

We first present in Fig.~\ref{fig:XPES} the ground-state and lowest-energy excited-state PESs of the proton desorption reaction.
There are a number of notable properties of the excited-state surface. Most strikingly, the activation barrier in the excited state, 2.64~eV, is \textit{less than half that of the ground-state desorption energy}, 5.98 eV.
Additionally, the attractive region where the proton experiences a restoring force towards the surface is significantly reduced from at least 3.0~\AA{} in the ground state to only 1.6~\AA{} in the excited state. 
We expect that the lowering of the barrier and shortening of the attractive region of the potential will significantly enhance the desorption of the proton in the excited state. 
While it may be difficult to directly observe changes in the reaction rate experimentally on \textit{e.g.} epitaxial \tioo{} thin-films, one may probe this excited-state PES by over-exciting the system with high energy light~\cite{jin2020critical}, which provides the extra kinetic energy to the excited-state reaction to more easily overcome the 2.6 eV barrier in experiments.
Our calculations also predict that near equilibrium, the excited-state PES is \textit{identical} to the ground state. In particular, this implies the equilibrium position is unchanged in the excited state within our computational sampling and that there is no softening of the OH vibrational bond, in contrast to the well-documented effect of phonon softening of photo- or carrier-doped semiconductors~\cite{chakraborty2012symmetry, fukuda2005doping}. 
Hence, excited-state Raman and infrared-spectroscopy can rule out other mechanisms relative to the photodesorption process predicted here. We emphasize that these predictions about the bond softening and equilibrium position are only possible due to the fine resolution of the PESs afforded by large-scale MBPT calculations.

The properties discussed above are strongly favorable for supporting enhanced kinetics in the excited state reaction, but nearly all of them are also nontrivial to understand. This lack of a general understanding of the \textit{qualitative} behavior of the excited-state PES is part of the reason for the lack of a systematic way to understand photocatalytic reaction dynamics. In the next section, we discuss at length the qualitative and quantitative behavior of these PESs, and explain how the lower barrier, shortened attractive region, equilibrium position, and lack of bond softening naturally arise from a very general physical picture of the underlying reaction mechanism.

\subsection{Reaction Mechanism}
Here we present an analysis of the evolution of various quantities at both the ground- and excited-state levels -- including densities of states, quasiparticle and excitonic wavefunctions, and energy alignments -- to fully analyze the qualitative and quantitative behavior of the excited-state PES reported in Fig.~\ref{fig:XPES}. For completeness, we note that we understand the ground-state PES as a simple Lennard-Jones-like potential, and that it can be well-fit by a potential of this form. 

\begin{figure}[H]
    \centering
    \includegraphics[trim={0 -1.1cm 0 0}, clip, width=0.6\textwidth]{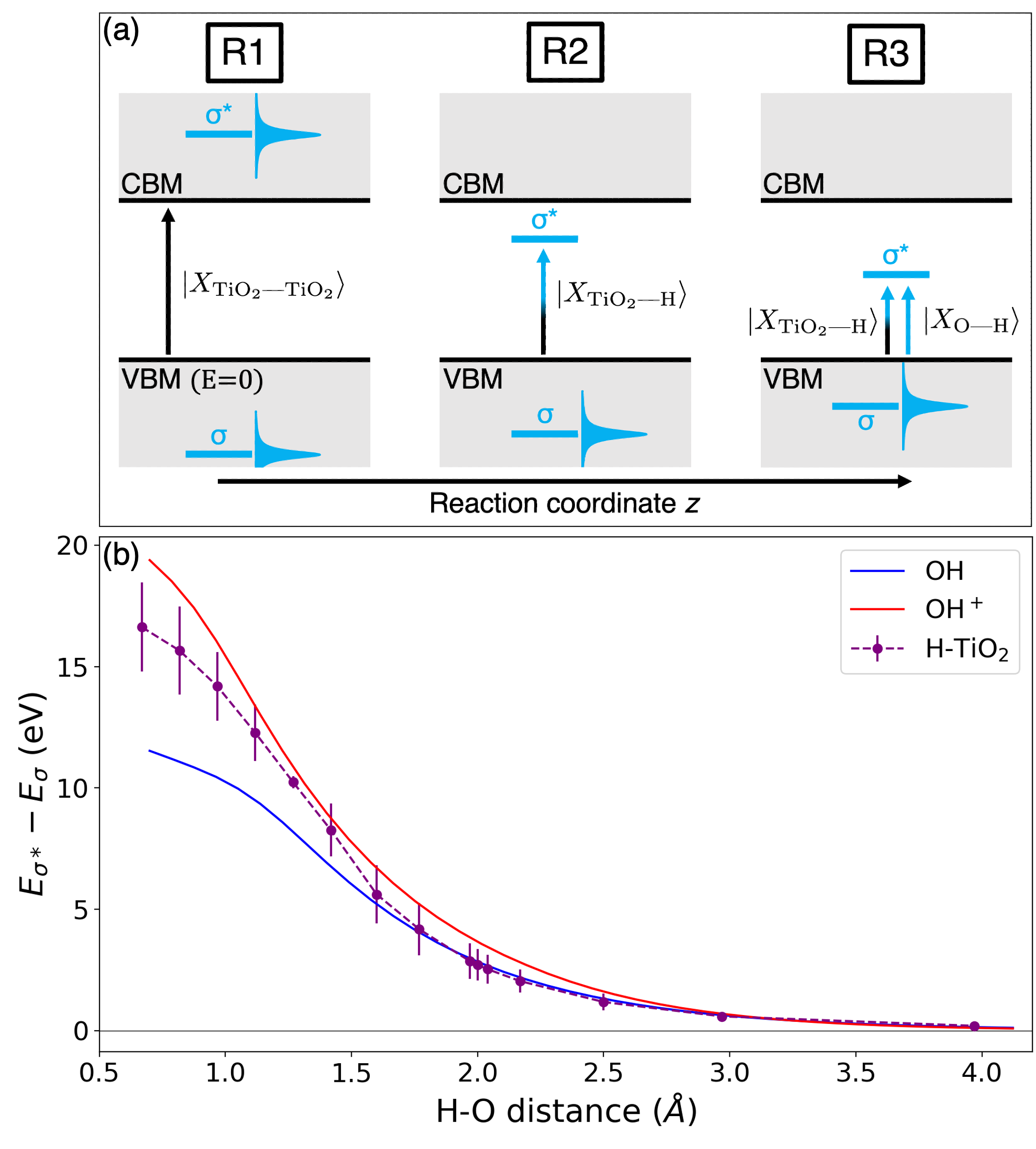}
    \caption{(a) Qualitatively distinct regions that occur along the reaction pathway, distinguished by the relative alignment of the \tioo\ manifolds and the bonding and antibonding orbitals of the molecular complex formed by the adsorbed proton and oxygen O$_{2c}$ on the \tioo\ surface. Lorentzians denote Fano resonances formed with the \tioo\ manifolds. Arrows and kets denote the types of transitions that form the lowest-energy excitons in the system in the three different regions. (b) Actual difference between antibonding and bonding energies at the DFT level, extracted from the projected density of states (pDOS) on the hydrogen orbitals (dashed curve). Error bars roughly represent the width of the Fano resonances. Also plotted is the same quantity for isolated molecules of OH and OH$^+$ (solid curves), showing large similarities with the behavior of the more complex catalytic system.}
    \label{fig:regions_dos-curve}
\end{figure}

Our physical picture involves two main subsystems.
The first subsystem contains the valence and conduction band manifolds of the \tioo{} slab which remains roughly unchanged throughout the desorption process. By restricting our discussion to a proton desorption process in the dilute limit and with much faster dynamics relative to the \tioo{} ions, one expects the surface geometry to stay roughly constant as a function of the reaction coordinate $z$.
In this limit, there should exist clear valence band maximum (VBM) and conduction band minimum (CBM) states from \tioo{} displaying quasiparticle excitation energies roughly constant relative to the vacuum level as a function of $z$ (Fig.~\ref{fig:regions_dos-curve}(a)). We explicitly confirm this identification by analyzing the density of states (DOS) evaluated with quasiparticle corrections (see Supporting Information).

The second subsystem contains two atomic-like orbitals from the adsorbed H$^{+}$ and the O$_{2c}$ on the \tioo{} surface.
These two atomic orbitals strongly interact, leading to bonding and anti-bonding OH-like molecular orbitals.
It is also clear that the two subsystems are strongly coupled, as one cannot isolate the relevant O$_{2c}$ atomic orbital from the \tioo{} bulk states.
Still, this coupling will be identified next using Fano resonance theory, which allows us to extract quantitative information from our first-principles calculations without requiring detailed information about the participating atomic-like orbitals.
Below we show how our analysis naturally identifies three distinct regions along the evolution of the reaction coordinate -- $R1$, $R2$, and $R3$ -- characterized by different alignments of various energy levels (Fig.~\ref{fig:regions_dos-curve}(a)), and explains the main features of the excited-state PES (Fig.~\ref{fig:XPES}).

To begin, as the O$_{2c}$-H separation $z$ increases, we expect the splitting between the resulting OH-like bonding and antibonding molecular orbitals to decay roughly exponentially and the molecular orbitals to evolve to the isolated atomic-like orbitals; for our system, the energies of the atomic-like orbitals are nearly degenerate as hydrogen and oxygen have very similar ionization potentials (IPs)~\cite{moore1949atomic}.
We find these atomic-like levels to lie below the VBM quasiparticle energy, consistent with the previously computed IP of \tioo~\cite{kashiwaya2018work}.
In the regions where the OH-like molecular orbitals are energetically overlapped with the \tioo{} manifolds, the discrete molecular orbitals evolve into resonances (Fig.~\ref{fig:regions_dos-curve}(a)).
Using a Fano formalism with typical assumptions (constant density of states on the continuum manifold and coupling matrix elements between the discrete and continuum states) one further expects these resonances to be described by Lorentzian distributions~\cite{Fano1961}. 
While it is impractical to compute the precise shape of these Fano resonances, their energy widths can be directly obtained from projected DOS (pDOS) of the quasiparticle wavefunctions (approximated here by Kohn-Sham states~\cite{hybertsen1986electron}) onto the orbitals of the adsorbed proton.
With this approach, we identify the center of the pDOS distributions below and above the gap with the energy of bonding and anti-bonding OH molecular orbitals, respectively, without any coupling to the continuum.
The energy difference between such anti-bonding and bonding orbitals in our full \tioo{} system is displayed as a function of $z$ as a dashed line in Fig.~\ref{fig:regions_dos-curve}(b), while the error bars represent the summed width of the resonances extracted from the pDOS (see Supporting Information) and is related to the hybridization of the individual bonding and antibonding orbitals with their respective continua. 
For comparison, we also compute the same quantity for isolated molecules of OH and OH$^+$, which provide nearly perfect lower and upper bounds for the curves we obtain on \tioo{} and is consistent with the expected charge state of the OH complex on the \tioo{} having a value between zero and one.

Aside from understanding the mean-field energetics, the advantage of this Fano-based analysis is in rationalizing the evolution of the excited-state PES, as it allows us to even predict the \emph{character} of the excitations as a function of $z$. 
In general, optical excitations are described by excitons, which are coherent combinations of several electron-hole pairs. 
The lowest-energy exciton primarily involves the lowest-energy quasiparticle transitions across the gap, which we mark with arrows and kets in Fig.~\ref{fig:regions_dos-curve}(a).
Depending on the relative alignments of various QP states discussed, qualitatively different excitons will form, so it is critical to obtain accurate QP levels with predictive methods such as the \textit{ab initio} GW approach. 

In region $R1$, due to their large splitting ($>5$ eV), the bonding and antibonding orbitals are deep in the valence and conduction manifolds, and the lowest-energy excitons are purely that of the \tioo. In this region, the optical excitation energy is nearly constant, and hence the excited-state potential energy surface is expected to be parallel to the ground-state one in $R1$.
Now, because the IPs of H and O are lower than that of \tioo{}, the antibonding orbital will at some point traverse the gap from CBM to VBM and lower the optical excitation energy in the process.
In $R2$ this process begins, and the antibonding orbital enters the \tioo{} gap and acts as an unoccupied defect state. 
The lowest energy quasiparticle transitions in this case are those between the VBM and the antibonding orbital. Hence, the electron participating in the lowest-energy exciton should move from the bulk of the \tioo{} to the proton adsorbate as $z$ transverse from $R1$ to $R2$. 
$R3$ is subtly different from $R2$ as the antibonding state is still in the gap, but here the Fano resonance formed by the bonding OH state and the valence manifold has moved up significantly in energy and approaches the VBM.
Critically, this results in the formation of a \textit{molecular-like exciton}, as it involves quasiparticle transitions not only from the \tioo{} VBM to the antibonding OH orbital previously found in $R2$, but now also from the bonding OH resonance to the antibonding OH orbital. These OH bonding-to-antibonding transitions are expected to increase the exciton binding energy, since they involve an electron-hole Coulomb interaction between quasiparticles that are spatially close. These two contributing transitions are shown with the two kets in Fig.~\ref{fig:regions_dos-curve}(a). We confirm all these predicted changes in the nature of the lowest-energy optical excitations obtained from our simple model by plotting the exciton wavefunction, which we discuss later in Fig.~\ref{fig:exciton_DOS_isos}~(a).

\begin{figure}[H]
    \centering
    \includegraphics[width=.7\textwidth]{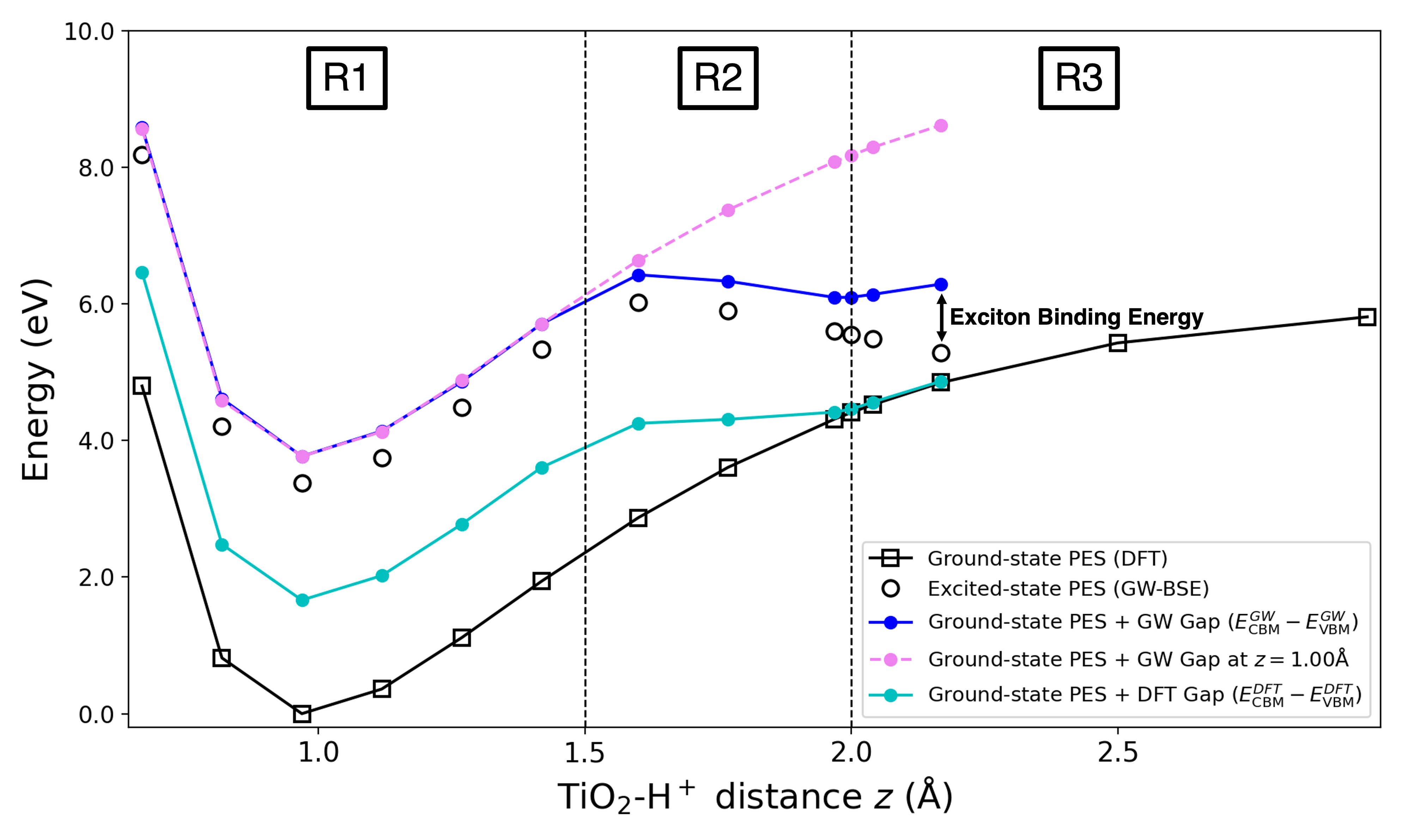}
    \caption{Comparison of qualitative and quantitative approximations to the true excited-state PES. The same data from Fig.~\ref{fig:XPES} for the ground-state and excited-state PESs are shown in black. Other curves show approximations to the excited-state surface by adding to the ground-state PES the GW quasiparticle bandgap at every point (dark blue), the GW quasiparticle bandgap held constant to its value at equilibrium (pink dashed), and the Kohn-Sham DFT bandgap at every point (light blue). The difference between the energies obtained by the dark blue curve and the black circles (\textit{i.e.}, the lowest-energy excited state obtained from the GW-BSE calculations), directly yields the exciton binding energy.}
    \label{fig:xpes_allgaps}
\end{figure}

\begin{figure}[H]
    \centering
    \includegraphics[width=.7\textwidth]{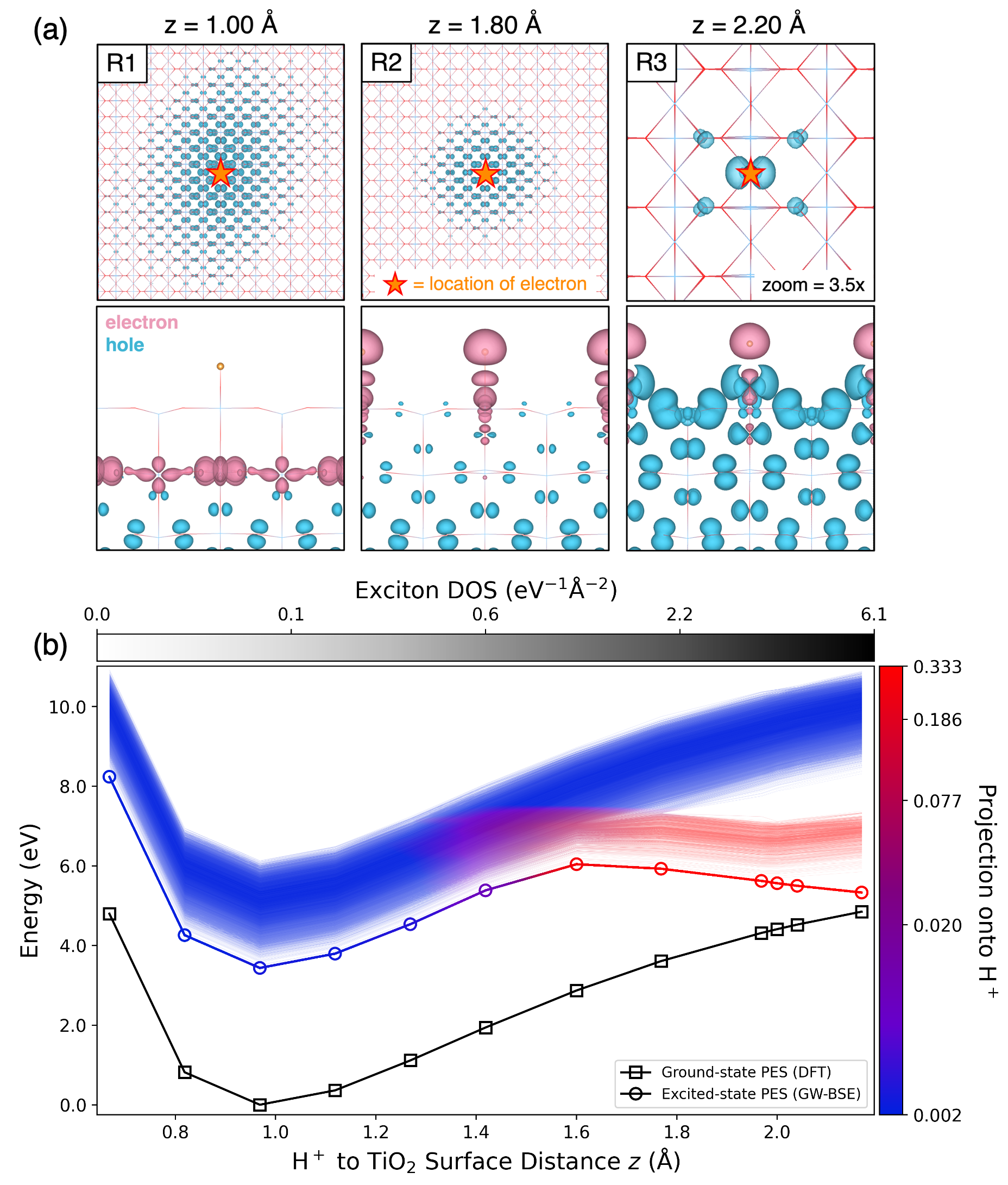}
    \caption{(a, top) Isosurfaces showing the exciton wavefunction with a fixed electron coordinate at the proton for each of the three regions. The isosurfaces shown enclose 0.5 of the hole density. (a, bottom) Isosurfaces of the excited-state charge density $\rho_S(\vb{r}) = \int d\vb{r}_h |\Psi_S(\vb{r}, \vb{r}_h)|^2 - \int d\vb{r}_e |\Psi_S(\vb{r}_e, \vb{r})|^2$ where $\Psi_S(\vb{r}_e, \vb{r}_h)$ is the exciton wavefunction. The positive and negative isosurfaces individually enclose 0.75 of the corresponding particle density. (b) Exciton DOS (opacity, top colorbar) and real-space projection onto the H$^+$ adsorbate (color, side colorbar) superposed on the excited-state PESs. The contribution of the hydrogen evolves smoothly from high to low energies as the reaction progresses, providing explicit evidence for the Fano resonance-based picture presented earlier.}
    \label{fig:exciton_DOS_isos}
\end{figure}

Aside from the expected exciton characters, we can now understand several features of the original excited-state PES in Fig.~\ref{fig:XPES}. 
In particular, due to the pure \tioo-character of the excitons in $R1$, the excited-state PES is simply a perfect replica of the ground-state PES, but offset by the lowest excitation energy, in agreement with the behavior found in Fig.~\ref{fig:XPES}. 
We show this explicitly by comparing the excited-state PES approximated by the sum of the ground-state PES and GW gap at equilibrium with the pink curve of  Fig.~\ref{fig:xpes_allgaps}, where it is clear that in $R1$ there is excellent agreement between the true excited-state surface and this approximation, up to the shift of the exciton binding energy that is constant to within 30 meV in $R1$. 
The fact that initial excitation is purely that of the \tioo\ has also been noted in other studies of different photocatalytic reactions on \tioo~\cite{jin2020critical, zhang2022interfacial, wang2024excitonic}, but these studies could not explicitly show the excited-state PES was simply a replica of the ground state due to the coarse sampling of the reaction coordinate.

Despite this, the entering of the antibonding orbital into the gap in $R2$ is the main feature of this reaction that yields the pronounced drop in the excited-state PES, and is the reason for the significantly reduced activation barrier and attractive region of the excited-state PES. This is almost entirely a result of the particular alignment of the various IPs, VBM, and CBM relative to vacuum, and can serve as a highly tunable knob in the design of different reactions, as we elaborate in the Discussion section below. 
An important conclusion of this finding is that, if not carefully engineered, \textit{there is not necessarily a lower activation barrier in the excited state}.
Furthermore, the formation of the molecular-like excitation in regions $R2$ and especially $R3$ dramatically localizes the corresponding excitonic wavefunctions and also significantly \textit{increases} the binding energy, from about 0.4~eV in $R1$ and $0.5$~eV in $R2$ to 1.0~eV in $R3$.
This surprising effect is a result of the very rapid change in the exciton character to molecular-like over a short distance, and rebuts the naive expectation that the exciton binding energy should decrease with increasing distance between the H$^+$ and \tioo.
This sharp increase in binding energy also greatly smooths the transition between the excited- and ground-state PESs, and makes it possible for a near-zero-energy nonadiabatic transition to occur after about $z=2.20$~\AA. 
In particular, it is notable that the neglect of excitonic effects severely overestimates the energy difference between the excited-state and ground-state PESs beyond $z\sim 2.00$~\AA.
These effects are visible in the blue curve of Fig.~\ref{fig:xpes_allgaps} which approximates the true excited-state PES with the sum of the ground-state PES and the GW quasiparticle gap at each point in the reaction coordinate. The difference between this curve and the true excited-state PES is the exciton binding energy.

The analysis of excitonic character based on the lowest energy transitions provides very useful heuristics for understanding the excited-state PES, which we now more rigorously discuss by explicitly analyzing the excitonic energies and wavefunctions. 
In particular, we show now that the most complete description of the reaction dynamics is obtained through analysis of the \textit{entire low energy excitation spectrum}, and  present excitonic isosurfaces for representative values of the reaction coordinate $z$ in each of the three regions. 
The bottom panel in Fig.~\ref{fig:exciton_DOS_isos}(a) shows the change in the total charge density upon exciting the system to the lowest-energy exciton, $\rho_S(\vb{r}) = \int d\vb{r}_h |\Psi_S(\vb{r}, \vb{r}_h)|^2 - \int d\vb{r}_e |\Psi_S(\vb{r}_e, \vb{r})|^2$, where $\Psi_S(\vb{r}_e, \vb{r}_h)$ is the exciton wavefunction, and explicitly confirms the expected evolution of the exciton character in each of the three regions. 
The transition from $R1$ to $R2$ is seen in the evolution of the electron density from the bulk of the \tioo{} to the adsorbed proton, and the transition from $R2$ to $R3$ is visible in the significant accumulation of hole density around the O$_{2c}$, corresponding to the strong molecular character of the exciton. 
Complementarily, we compute the isosurfaces of the exciton wavefunction for the electron fixed at the proton position, $|\Psi_S(\vb{r}_h, \vb{r}_e=\mathrm{const.})|^2$, in the top panel of Fig.~\ref{fig:exciton_DOS_isos}(a). 
The dramatic shrinking of the Bohr radius for the lowest-energy exciton is quite visible as we transition from $R1$ to $R3$ and, as mentioned previously, accompanies the increase in the exciton binding energy. 

The physical picture presented in Fig.~\ref{fig:regions_dos-curve}(a) can be extended to understand the full excitation spectrum, and not just the lowest-energy state. 
The evolution of the two Fano resonances formed by the bonding and antibonding states should be visible not only in the pDOS of the single-particle orbitals as previously shown in Fig.~\ref{fig:regions_dos-curve}(b), but also in the character of all excitons in the excitation spectrum.
We see this explicitly by performing real-space projections of all the exciton wavefunctions onto the proton adsorbate, shown in Fig.~\ref{fig:exciton_DOS_isos}(b) along with the exciton DOS.
At large $z$, there are two distinct branches of excited states -- one containing excitons displaying a large contribution from the adsorbate, and another branch without such contributions.
Moreover, the projection smoothly fades and broadens at smaller values of $z$, a result of the broader Fano resonances formed and the hybridization of different transitions induced by the electron-hole interaction.
This is a direct confirmation of the physical analysis presented in this work.

For completeness, we note that the projection values $P_S$ onto the proton for each exciton $\ket{S}$ can be defined in terms of the exciton expansion coefficients $A^S_{vc\vb{k}}$ obtained from solving the BSE, $\ket{S} = \sum_{vc\vb{k}} A^S_{vc\vb{k}} c_{v\vb{k}} c^{\dagger}_{c\vb{k}} \ket{0}$, where $c_{n\vb{k}}$ is a quasiparticle destruction operator at band $n$ and wavevector $\vb{k}$, and $\ket{0}$ is the electronic ground state.
From the solutions of the BSE, we evaluate $P_S = \sum_{vc\vb{k}} |A_{vc\vb{k}}^S|^2 P_{vc\vb{k}}$, where $P_{vc\vb{k}} = \frac{1}{2}(P_{v\vb{k}} + P_{c\vb{k}})$ is the average of the quasiparticle projections onto the H$^+$.
We define $P_{n\vb{k}}$ as the fraction of the corresponding real-space charge density $|\psi_{n\vb{k}}(r)|^2$ associated with the quasiparticle state at $n\vb{k}$ above the horizontal plane halfway between the H$^+$ and O$_{2c}$ (see Fig.~\ref{fig:struct}). 
We emphasize that this kind of analysis is uniquely suited to MBPT, as it is straightforward to obtain large numbers of excited state wavefunctions.

\subsection{Discussion}
The favorable properties of the excited-state reaction studied here are partially a result of a fortunate choice of the particular reactant, catalyst, and active site in terms of the relative alignments of the IPs and semiconductor manifolds relative to the vacuum level.
It is easy to imagine that the same analysis could be carried out for another similar reaction but with a different adsorbate-catalyst combination whose energy alignments do not yield a reduced reaction barrier in the excited state.
Indeed, even from the single case study performed here, it is already possible to extract several general design principles that can guide the design of future photocatalytic reactions based on tabulated energy levels.
As mentioned in the results, the main reason for a reduced activation barrier in the excited state for the present system is that the antibonding state happened to traverse the gap of the catalyst as $z$ increased. 
In general, the antibonding-bonding splitting of the atomic states participating in the reaction will be very large near equilibrium, so \textit{it is unlikely to see differences in the excited-state PES near equilibrium}, at least for single- or few-atom adsorbates. 
It is possible to engineer the IPs of the relevant atoms relative to the catalyst manifolds to ensure that either the bonding or antibonding state crosses the gap and hence decreases the excited-state barrier. 
In particular, since the IPs of the reactant and active site atoms are the energies that the bonding and antibonding states evolve to as the desorption reaction progresses, it is sufficient to choose the higher of the two IPs to lie below the CBM (preferably far below) or the lower of the two IPs to lie above the VBM (preferably high above). 
This ensures that either the antibonding state traverses the gap from above or the bonding state traverses the gap from below, respectively. To ensure a full crossing of the gap, the relevant IP should be at least the band gap away from the relevant manifold extremum.
Additionally, it is beneficial to choose a catalyst whose gap is similar to or larger than the ground-state activation barrier. 
Even if there is a state that crosses the gap as the reaction progresses, if the excitation energy is too small, no reduction in the activation barrier will occur. 
This is what occurs for the system studied here if GW corrections are not applied, due to the significant underestimation of the gap by DFT, as shown by the light blue curve in Fig.~\ref{fig:xpes_allgaps}. 
Thus, simply by considering the band gap, IP of the catalyst, and IPs of the reactant and active site, one can engineer a good starting point for the design of an efficient photocatalyst for a particular reaction. 

Beyond simple consideration of the alignment of these critical energies, our Fano resonance-based analysis allows explicit estimates of the true excited-state PES based just on DFT calculations of the elementary reaction steps. This approximated excited-state PES is much cheaper to compute, and can yield qualitative predictions for the excited-state PES using only the ground-state PES and experimental IPs as input – suitable for subsequent high-throughput investigations. We present some details on how to construct this approximate PES in the Supporting Information.

We also note how defects can be accounted for within our framework, such as the common oxygen vacancy in \tioo~\cite{nakamura2000role, li2017synergistic, bi2021tuning, zhao2015effect, chen2018facile, feng2018significantly, qi2018photocatalytic}. While the desorption of chemical species attached to vacancies requires specific calculations to capture the local bonding environment, atomic reconstructions, and the level alignment of defect states, the effect of nearby vacancies can be easily estimated. For instance, the oxygen vacancy in \tioo\ introduces in-gap unoccupied quasiparticle states~\cite{morgan2009density} that lower exciton energies, reducing the desorption activation energy. This provides additional generalizability for our conclusions, and the underlying Fano model, to materials with defects.

\section{Conclusions}
We present a rigorous calculation and analysis of the excited-state potential energy surface for the desorption of a proton from rutile \tioo\ (110) computed with the \textit{ab initio} GW and GW-BSE approaches. 
We find a significant reduction in the activation barrier upon excitation and no bond softening.
We provide a detailed qualitative and quantitative analysis of the observed behavior, with an underlying physical picture based on the formation and evolution of Fano resonances formed by molecular orbitals of the adsorbate and active site in the presence of the \tioo\ manifolds. 
These resonances naturally partition the excited-state PES into three regions that host qualitatively different excitons, which transition from catalyst-like to adsorbate-like, and experience strong localization and a significant increase in binding energy even as the adsorbate is pulled away from the catalyst surface. 
Finally, we show that the most complete description of the excited-state reaction is obtained by analyzing the excitonic wavefunctions of the entire excitation spectrum and not just the lowest energy exciton, which is uniquely accessible with MBPT. 

The physical understanding presented here is generalizable and will be significantly extended in a future work. 
It is also amenable to data-driven approaches to generate more design principles tailored to specific classes of photocatalytic reactions. We anticipate being able to predict which materials are good photocatalysts for a given reaction, beyond simple desorption reactions.

\section{Methods}
\subsection{Density-Functional Theory}
To accurately treat the slab and hydrogen bonds, we employ the Perdew-Burke-Ernzerhof exchange-correlation functional~\cite{perdew1996generalized} with van der Waals~\cite{thonhauser2015spin, thonhauser2007van, berland2015van, langreth2009density} and dipole corrections in the Quantum ESPRESSO code~\cite{QE-2017}. 
A $4\times 4\times 1$ Monkhorst-Pack $\vb{k}$-point grid and 80 Ry and 320 Ry wavefunction and charge density planewave cutoffs are used, respectively, in our calculations. A $0.001$~Ry Gaussian smearing of the occupations was used to aid convergence.
We do not employ the common Hubbard correction via the DFT+U method~\cite{morgan2007dft+, yan2013band}, as this is a semi-empirical approximation to the quasiparticle self energy, which we compute from first principles within the GW method. All calculations were spin-unpolarized as spin degeneracy is maintained until after the crossing of the ground and excited-state PESs, and an extended discussion is provided in the Supporting Information. All calculations (including MBPT calculations) were performed at 0K, as temperature effects are negligible due to the large electronic energy scales involved. All crystal structures are reported in the Supporting Information.

\subsection{Many-Body Perturbation Theory}
We utilize the BerkeleyGW code~\cite{deslippe2012berkeleygw} for GW and BSE calculations. We compute quasiparticle properties at the single-shot G$_0$W$_0$ level using a dielectric cutoff of 20 Ry and the Hybertsen-Louie generalized plasmon-pole model~\cite{hybertsen1986electron}, with significant acceleration and complete convergence with respect to the number of bands in the calculations provided by our recent mixed stochastic-deterministic approach~\cite{altman2024mixed}. The use of our stochastic approach reduces the scaling of the MBPT approach from quartic to sub-cubic for typical system sizes such as those studied here, and we refer the reader to the references cited in this section for an extended discussion of the scaling of the approach. Coulomb truncation in the out-of-plane direction is not used to better reproduce the dielectric environment of a semi-infinite surface (see Supporting Information). We employ the Rohlfing and Louie interpolation scheme to evaluate the BSE on a $8 \times 8 \times 1$ $\vb{k}$-point grid~\cite{rohlfing2000electron}, with 11 valence and 11 conduction states to converge the lowest-energy excitation.

\section*{Acknowledgement}
A.R.A. and F.H.J. acknowledge helpful discussions from S. Kundu, M. Del Ben, and D. F. Swearer. 
This work was primarily supported by the Center for Non-Perturbative Studies of Functional Materials Under Non-Equilibrium Conditions (NPNEQ), funded by the US Department of Energy (DOE) Office of Science under contract DE-AC52-07NA27344. The work on the catalytic model was supported by the Keck Foundation (Grant No. 994816). This work uses codes developed by the Center for Computational Study of Excited State Phenomena in Energy Materials (C2SEPEM), which is funded by the DOE Office of Science under Contract No. DE-AC02-05CH11231. This research used computational resources at the Oak Ridge Leadership Computing Facility, a DOE Office of Science User Facility supported under Contract DE-AC05-00OR22725. Real-space and convergence analysis were performed using computational resources of the National Energy Research Scientific Computing Center (NERSC), a U.S. DOE Office of Science User Facility located at LBL, operated under Contract No. DE-AC02-05CH11231.

\section*{Supporting Information Available}
Extended computational and convergence details, relaxation and spin polarization effects, details on the extraction of Fano resonances, construction of heuristic potential energy surfaces, and all crystal structures.

\printbibliography[heading=bibintoc, title={References}]
\end{refsection}

\clearpage
\appendix
\setcounter{figure}{0}
\renewcommand{\thefigure}{S\arabic{figure}}
\begin{refsection}
\section*{\centering Supporting Information for \textit{Ab Initio} Mechanisms and Design Principles for Photodesorption from TiO${}_2$}
\section{Computational Details}\label{compdeets}
All density-functional theory (DFT) calculations were performed with the Quantum ESPRESSO plane-wave code~\cite{QE-2017} with the Perdew-Burke-Ernzerhof (PBE) exchange-correlation functional~\cite{perdew1996generalized}. We employed scalar-relativistic norm-conserving pseudopotentials from the PseudoDojo library~\cite{van2018pseudodojo}. GW and GW-plus Bethe-Salpeter equation (GW-BSE) calculations were performed in the BerkeleyGW code~\cite{deslippe2012berkeleygw}. GW calculations utilized the single-shot G$_0$W$_0$ approach and the Hybertsen-Louie generalized plasmon pole model for the frequency dependence of the dielectric function~\cite{hybertsen1986electron}. Significant acceleration and full convergence with respect to the sums-over-bands was provided by our recent algorithmic development on a mixed stochastic-deterministic approach to GW calculations~\cite{altman2024mixed}. BSE calculations were performed with the Tamm-Dancoff approximation and with interpolation from coarse to fine $\vb{k}$-point grids for proper convergence. 

Crystal structures were constructed with the Python package \texttt{ase}~\cite{larsen2017atomic} following the rutile \tioo{}(110) structures found in the literature~\cite{Tritsaris2014dynamics}.
The vacuum is chosen to be approximately equal in thickness to the slab to ensure periodic images in the out-of-plane direction cannot interact and capture the correct dielectric environment (see section~S\ref{convergence} below). 
A $2\times 1$ in-plane supercell is chosen to ensure small interactions between the adsorbed protons of adjacent cells (see center subplot of Fig~\ref{fig:spin-bands}). The equilibrium structure with the proton adsorbate is relaxed and the structure of the \tioo{} is kept fixed for the other calculations away from equilibrium (see section~S\ref{relaxation_effects} below for an extended discussion). 
To simulate the proton and maintain the semiconducting character of the system, we study the charged system with one electron removed; at the DFT level, we add a compensating jellium background to maintain charge neutrality. 
We also include van der Waals corrections in the form of the vdW-df-c09 exchange-correlation functional~\cite{thonhauser2007van, berland2015van} as implemented in the Quantum ESPRESSO package. 
Additionally, we include dipole corrections to counteract any spurious dipole moment formed due to the finite slab. 
We add a small electronic smearing ($\sim$10 meV) to help aid convergence. 

\subsection{Convergence with $\vb{k}$-points and Energy Cutoffs}\label{convergence}
We discuss convergence tests over the three main convergence parameters in the calculation: the $\vb{k}$-point grid, wavefunction cutoff, and dielectric cutoff. 
We choose a $\vb{k}$-grid of $4\times 4\times 1$ for the mean-field calculation and GW calculations of the dielectric function and quasiparticle (QP) energies. 
We also choose an 80 Ry plane-wave wavefunction cutoff. 
We find a change in the DFT total energy of 200 meV between two calculations with an 80 Ry wavefunction cutoff at $\vb{k}$-grids of $2\times 2\times 1$ and $4\times 4\times 1$, and similarly find a roughly 200 meV energy difference between calculations at 80 Ry and 100 Ry wavefunction cutoffs with a $4\times 4\times 1$ $\vb{k}$-grid. We consider these to be sufficiently converged as a result, and these convergence parameters are similar to those reported in other MBPT studies of \tioo~\cite{chiodo2010self, kang2010quasiparticle, landmann2012electronic}. As noted in the previous section, our new mixed stochastic-deterministic approach for GW calculations~\cite{altman2024mixed} alleviates the need to perform convergence tests on band truncation parameters.

\begin{figure}[H]
    \centering
    \begin{subfigure}[b]{0.49\textwidth}
        \centering
        \includegraphics[width=\textwidth, trim={1.5cm 0cm 2.5cm 1cm}, clip]{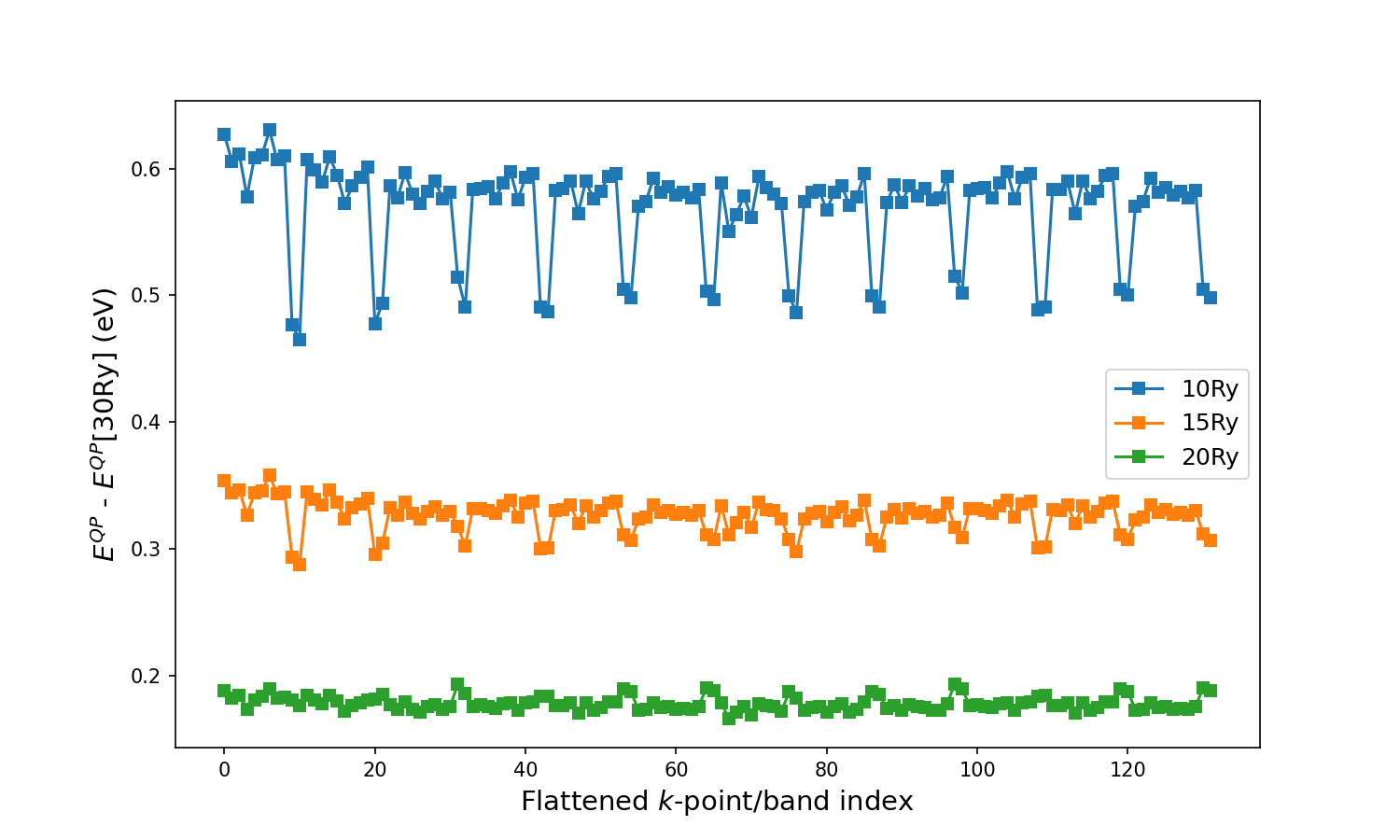}
        \caption{}
        \label{fig:eqp_vs_epscut}
    \end{subfigure}
    \hfill    
    \begin{subfigure}[b]{0.49\textwidth}
        \centering
        \includegraphics[trim={1.5cm 0cm 2.5cm 1cm}, clip, width=\textwidth]{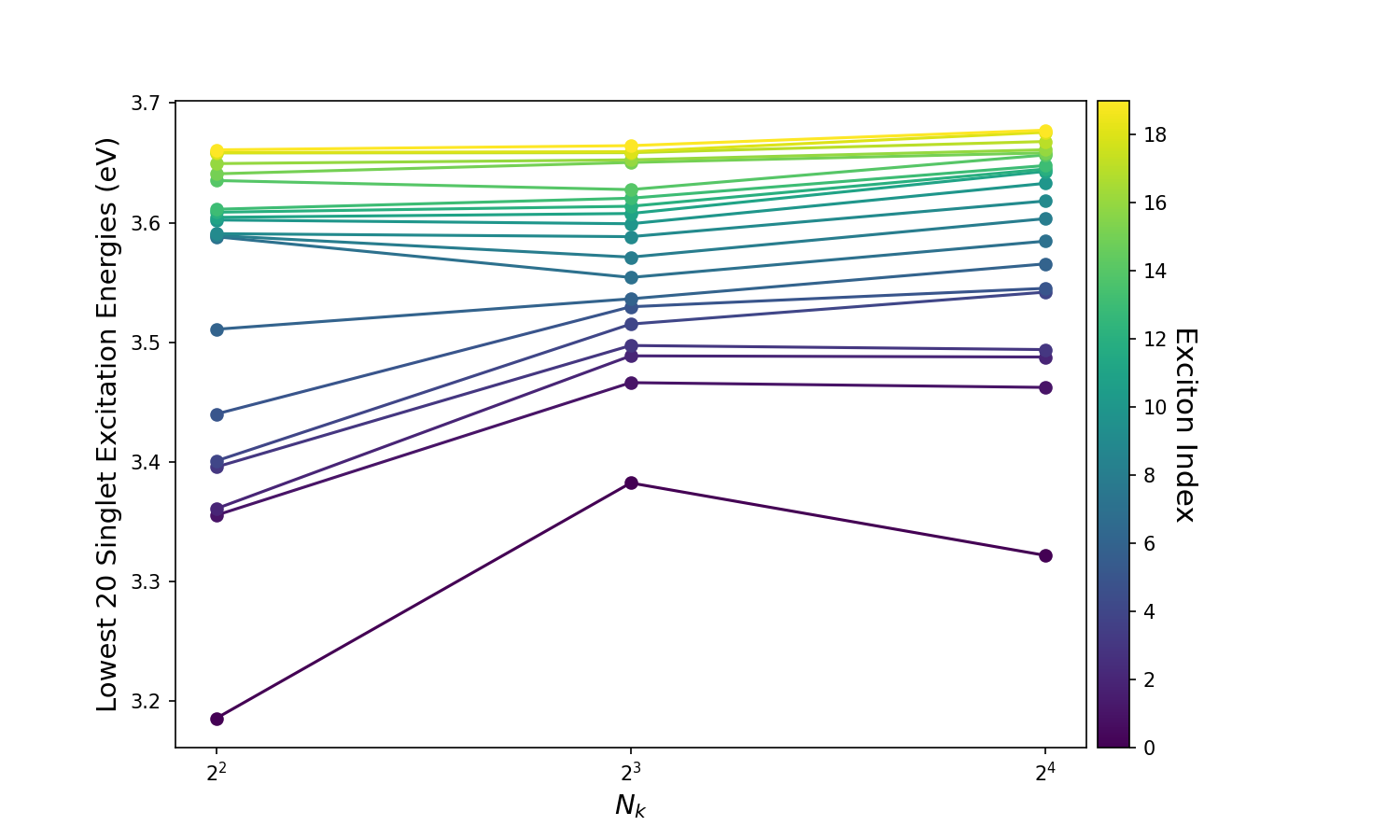}
        \caption{}
        \label{fig:bse-kgrid-conv}
    \end{subfigure}
    \caption{(a) Convergence of the quasiparticle energies vs dielectric cutoff. A large overall shift is visible across all states, but various band-to-band transition energies are much more tightly bounded, as the variance of each curve is much smaller than its average value. States shown include 9 valence bands and 2 conduction bands for the 12 unique $\vb{k}$-points in the coarse grid. (b) Convergence of the lowest 20 exciton energies with $\vb{k}$-point sampling. Both plots are for the equilibrium geometry.}
    \label{fig:GW-BSE_conv}
\end{figure}

In the calculation of the GW self-energy, we employ a 20 Ry cutoff for the screened Coulomb interaction and an 80 Ry cutoff for the bare Coulomb interaction, and we report a direct quasiparticle band gap of 3.77 eV at the $\Gamma$ point for the equilibrium structure, in good agreement with literature~\cite{chiodo2010self, kang2010quasiparticle, landmann2012electronic}. 
Our choice of a 20 Ry dielectric cutoff results in a nearly constant 200 meV shift in QP energies from a more converged calculation employing a 30 Ry cutoff (Fig~\ref{fig:eqp_vs_epscut}), but the band gap is converged to within 5 meV with the 20 Ry cutoff.
To model the dielectric environment of a semi-infinite slab, we do not employ Coulomb truncation in the out-of-plane direction. 
This results in no sharp feature in the dielectric function as is found in true quasi-2D materials, and partially alleviates the requirement of an extremely fine $\vb{k}$-point sampling~\cite{Qiu2016,Jornada2017}. 
Additionally, near the surface of a true semi-infinite slab, one roughly expects the dielectric constant to take a value of $\frac{1}{2}(1 + \varepsilon_{\text{bulk}})$, and for the slab calculation we actually compute, we expect the dielectric constant to be roughly $r\cdot (1 + \varepsilon_{\text{bulk}})$, where $r$ is the fraction of the unit cell that the slab takes up. For our system $r$ is about $0.4$. 
In our calculations we find this estimate to be qualitatively correct, as our long-range static dielectric constant evaluates to about $\varepsilon_{00}(\vb{q}\rightarrow 0, \omega=0) \sim 3.3$ at the equilibrium structure, consistent with the the value of the bulk dielectric constant in previous calculations on bulk rutile \tioo\ of $7-8$~\cite{kang2010quasiparticle, vast2002local} (here the $00$ subscript denotes the $\vb{G}=\vb{G'}=\vb{0}$ plane-wave components, $\vb{q}$ is the wavevector, and $\omega$ is the frequency).

We construct the BSE with 11 valence and 11 conduction bands, enough to converge the lowest 200 excitons to within 50 meV as compared to a calculation with only 5 valence and 5 conduction bands. 
We obtain a lowest singlet excitation energy of 3.32 eV at the equilibrium structure, in good agreement with literature values~\cite{chiodo2010self, kang2010quasiparticle, landmann2012electronic}. 
We find that we need a $\vb{k}$-point grid of $8\times 8 \times 1$ to converge the BSE to within $\sim$50 meV (Fig~\ref{fig:bse-kgrid-conv}), which is accomplished \textit{via} the interpolation scheme in BerkeleyGW. 

\subsection{Effect of Relaxation}\label{relaxation_effects}
\begin{figure}[H]
    \centering
    \includegraphics[width=.75\textwidth, trim={0 0 0 0},clip]{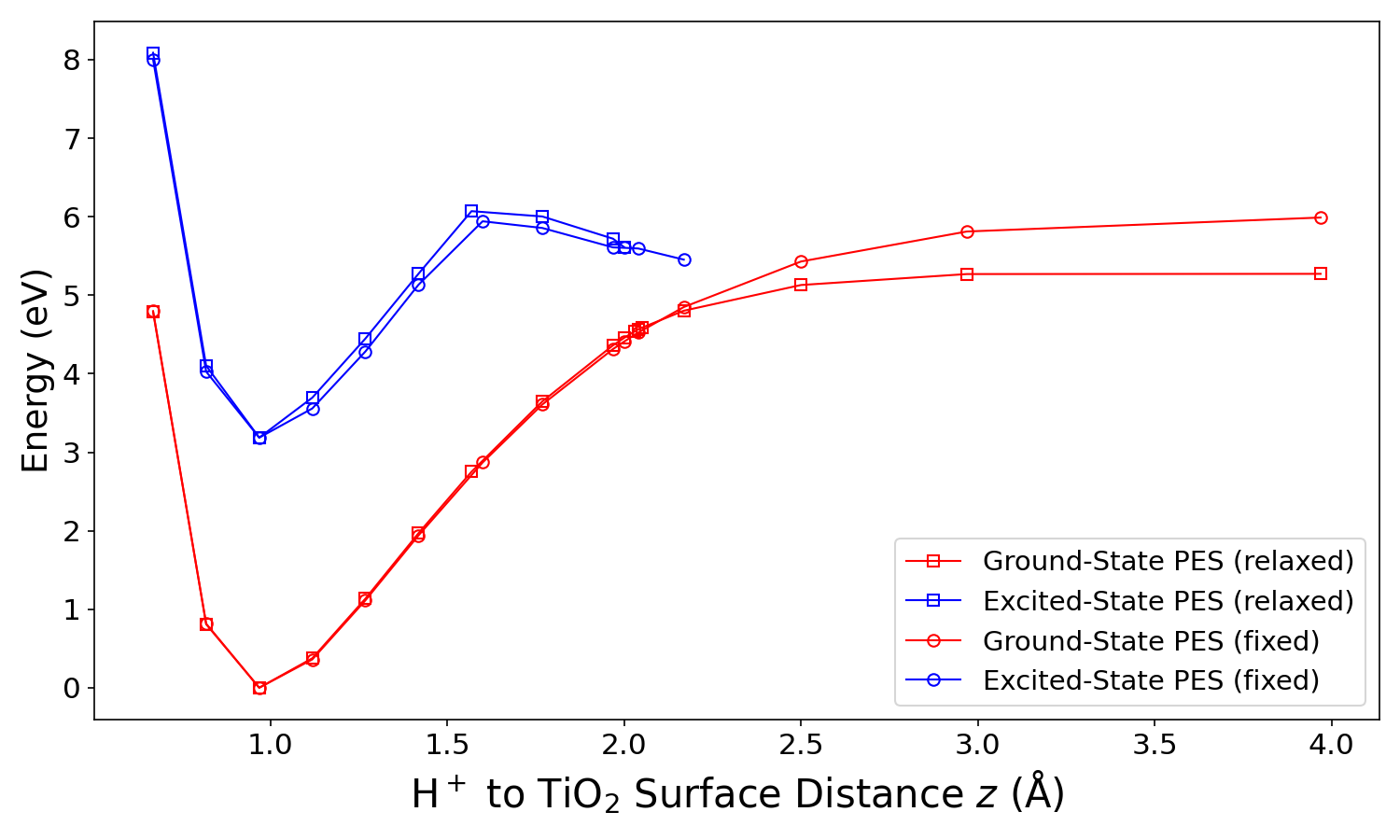}
    \caption{Excited-state PES computed with and without relaxation of the ground state structure.}
    \label{fig:relax_vs_fix}
\end{figure}

In Fig~\ref{fig:relax_vs_fix} we show the effect of relaxation of the ground state structure as opposed to the systems studied in the main text where the equilibrium structure for the \tioo is maintained at all distances.
To avoid extra computational expense, the calculations shown here use the slightly under-converged $\vb{k}$-point sampling $4\times 4\times 1$ in the BSE, with all other convergence parameters the same as written above. 
The structure at the equilibrium configuration is the same for both the fixed and relaxed PESs. There are negligible differences in both the ground and excited-state surfaces in the region before the crossing of the ground and excited states where the excited-state PES is well-defined, showing that relaxation effects are safe to ignore.

\begin{figure}[H]
    \centering
    \includegraphics[width=0.7\linewidth, trim={1cm 0 3cm 1cm},clip]{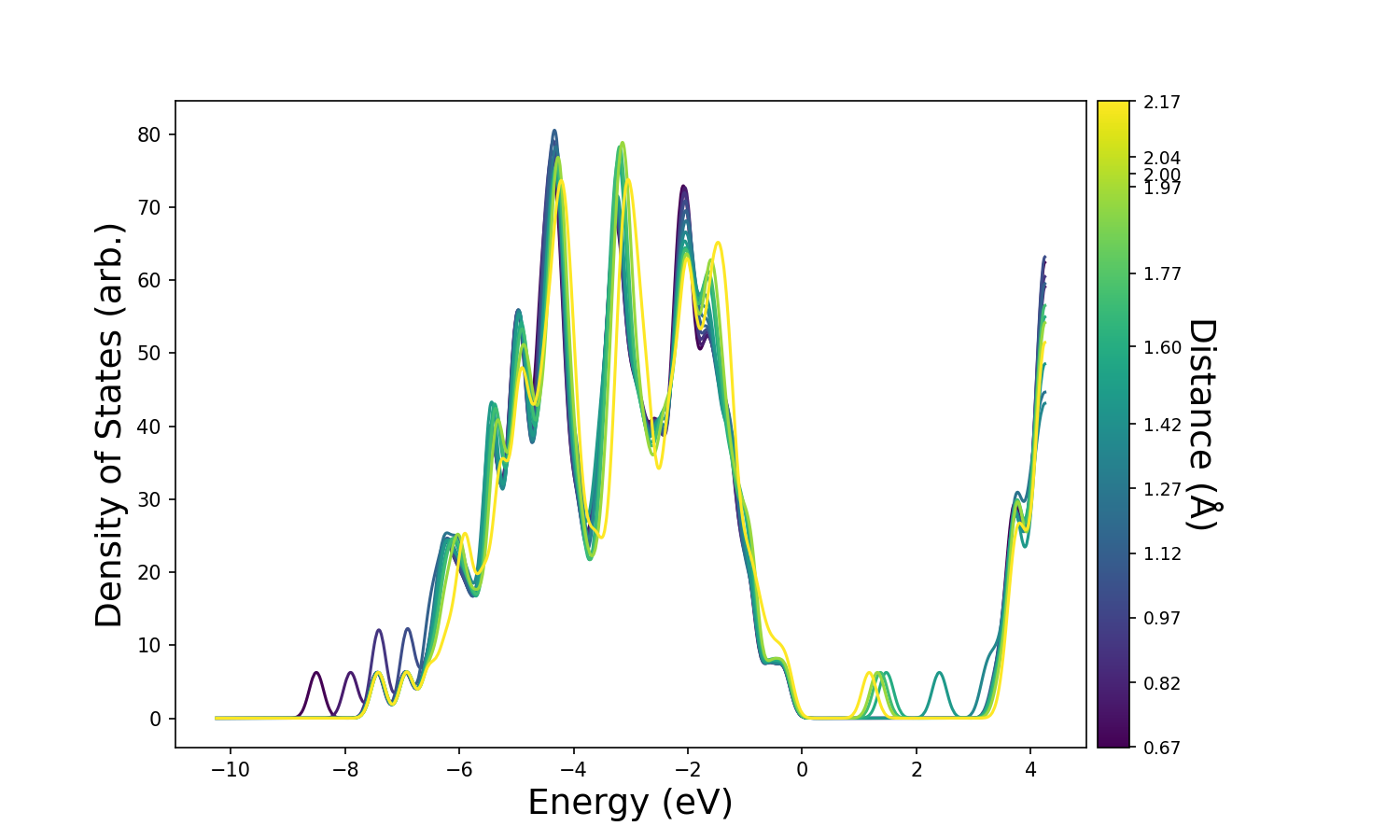}
    \caption{GW density of states for all sampled reaction coordinates, showing constant gap and valence manifold bandwidth. The energy scale is zeroed to the valence band maximum.}
    \label{fig:gw_dos}
\end{figure}

Additionally, in Fig~\ref{fig:gw_dos} we show the evolution of the GW density of states with the reaction coordinate. As we argue in the main text, in the dilute limit and due to the much faster dynamics of the proton as compared to the TiO$_2$ ions, structural deformations as the proton is desorbed are likely negligible, so we expect the valence and conduction manifolds to remain unchanged as the reaction progresses. It is shown explicitly in Fig~\ref{fig:gw_dos} that the bandwidth of the valence manifold and band gap is nearly constant as the reaction progresses, justifying our schematic of the constant valence and conduction band extrema in the main text. The small peaks visible at low energies at low distances are related to the evolution of the OH antibonding state (see main text and section~\ref{supp_fano}). 

\subsection{Effect of Spin Polarization}
The calculations of PES presented here are all performed within a scalar-relativistic, spin-unpolarized formalism. 
This is generally acceptable because the interesting dynamics of the system occur when the proton is strongly hybridized with the \tioo, and no magnetism is expected due to the closed-shell nature of the system. 
However, at large distances, we expect the proton to pick up one electron from the \tioo{} and for the system to take the form of two independent subsystems, TiO$_2^+$, and neutral H. 
In this case, we also expect the exchange interaction to split the energies of the states with respect to spin.
As such, spin-polarized calculations may be required before the isolated subsystems limit. 

\begin{figure}[H]
    \centering
    \includegraphics[width=\textwidth, trim={1cm 0 0 0},clip]{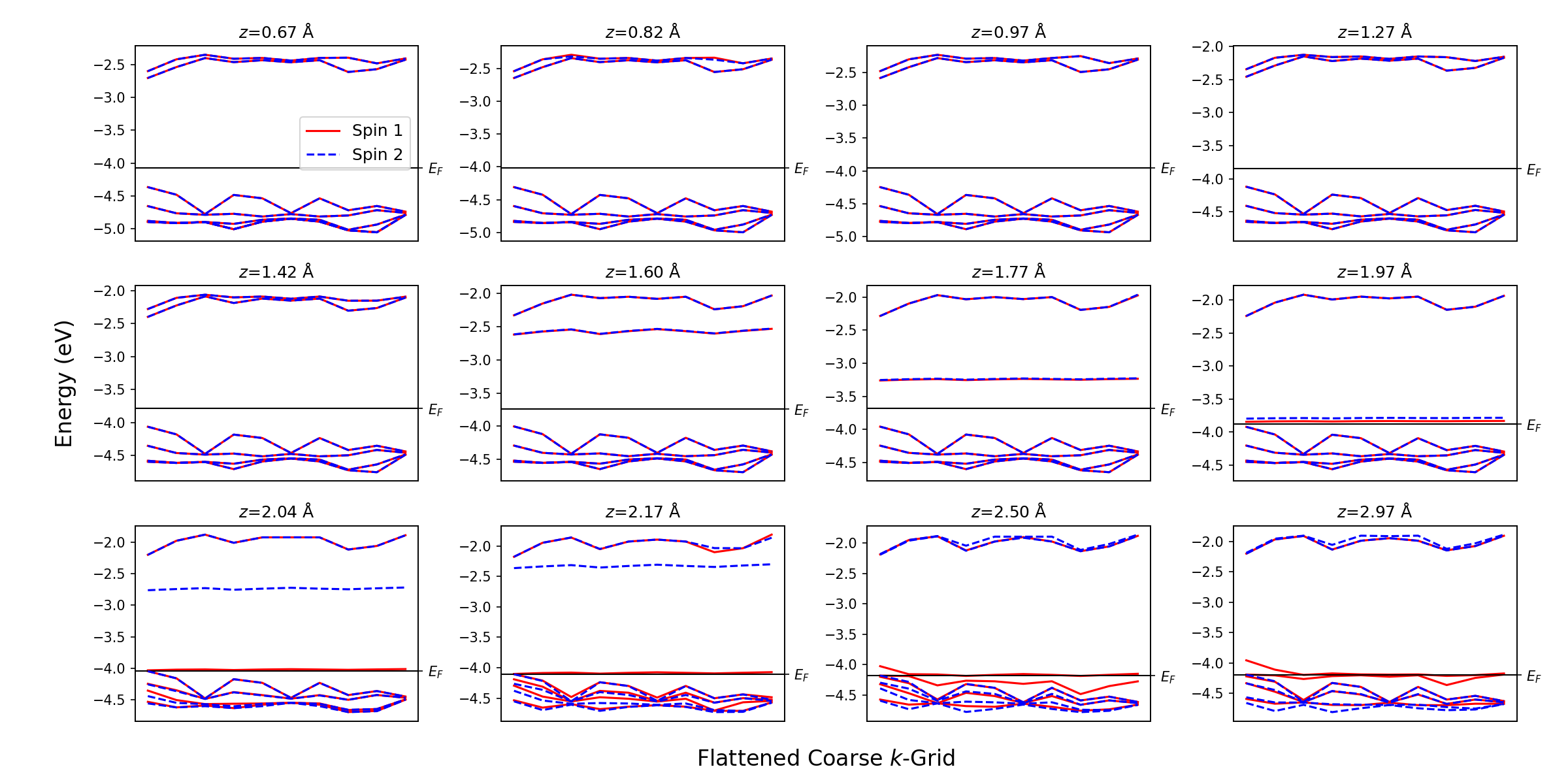}
    \caption{Spin-polarized DFT bands of fixed-geometry structures on the flattened regular $\vb{k}$-point grid. A significant exchange-driven spin splitting occurs only around $z=2.0$~\AA, which corresponds to the crossing of the excited-state and ground-state PESs and occupation of the hydrogenic state, as can be seen by the crossing of the Fermi energy (black line). Note that the labels ``Spin 1'' and ``Spin 2'' are arbitrary and do not correspond to a majority or minority spin configuration. Hence, they should not be compared across different values of $z$. Also, note that the system, within our supercell setup, becomes gapless after $z\sim 1.97$~\AA, after which the ground- and excited-state PES become degenerate.}
    \label{fig:spin-bands}
\end{figure}

In Fig~\ref{fig:spin-bands} we show spin-polarized DFT bands on the fixed-geometry structures, plotted against the flattened regular $\vb{k}$-point grid. 
The expected splitting occurs only near $z=2.0$~\AA, and corresponds with the crossing of the excited-and ground-state PESs. 
This splitting is a direct result of the crossing of the PESs, as the exchange split is dependent on the occupation of the hydrogen orbital, which only becomes significant near the crossing. 
As such, we see that our analysis based on spin-unpolarized calculations is valid for the region in which the excited-state PES is well-separated from the ground state, consistent with the adiabatic approximation used in our analysis. 
Moreover, near the crossing point around $z=2.0$~\AA, we see that one of the spins in the spin-polarized calculation follows the same behavior as the unpolarized calculation and closes the gap, so our results remain unchanged.

\section{Fano Resonances from Projected Density of States}\label{supp_fano}
As described in the main text, the energetic overlap of the bonding and antibonding states formed by the adsorbed proton and O$_{2c}$ from the \tioo\ with the valence and conduction band manifolds of the \tioo\ results in the formation of Fano resonances~\cite{Fano1961}, wherein the hybridized discrete-continuum state takes the form of a linear combination
\begin{align}
    \Psi_E &= a(E)\phi + \int dE'\; b_{E'}(E)\psi_{E'}. \label{fano_state}
\end{align}

Here $\Psi_E$ is the state resulting from the Fano resonance, $\phi$ is the discrete state, $\psi_{E'}$ is the continuum of states indexed by energy (assuming no degeneracies), and $a$ and $b_{E'}$ are coefficients. 
We recognize the coefficient $|a|^2$ as nothing more than the projected density of states (pDOS) of the discrete orbital, \textit{i.e.} the pDOS of the adsorbed proton (or O$_{2c}$). 
For simplicity, we focus on the pDOS of the adsorbed proton.

\begin{figure}[H]
    \centering
    \includegraphics[width=.9\textwidth, trim={0cm 0 0cm 0},clip]{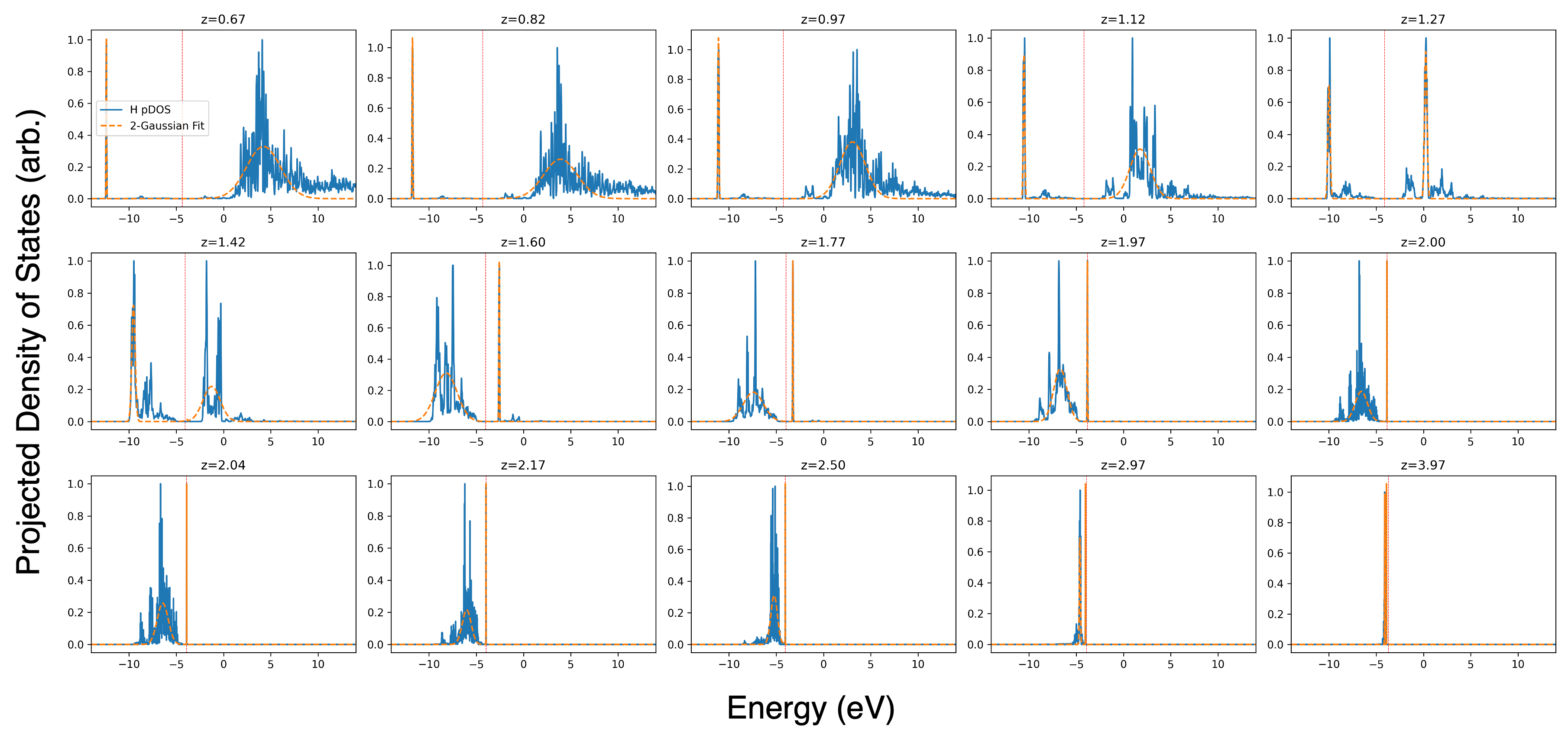}
    \caption{Two-gaussian fit to pDOS of the adsorbed proton. The horizontal axis denotes the single-particle Kohn-Sham energy, in eV, and the title of each plot is the distance $z$, in \AA{}. The pDOS is partitioned at the Fermi energy and normalized separately to have a maximum value of 1. Dashed red line denotes the \tioo\ VBM; note that the antibonding state dips below the VBM at large distances.}
    \label{fig:gaussian_fits}
\end{figure}

As described in the main text, Fano derived that in the simplest case the coefficient $|a(E)|^2$ takes the form of a Lorentzian function. In general, when the continuum states have a non-constant density of states, and/or the coupling matrix elements are not constant, we expect this term to still take the form of a peaked function, although with a different functional form; accordingly, in our analysis, we fit Gaussian curves to the pDOS as a proxy to a generic peaked function. The results of this fitting are shown in Fig~\ref{fig:gaussian_fits}, where the pDOS at each sampling of the reaction coordinate is fit with two Gaussian functions separately, corresponding to the two resonances formed with the valence and conduction manifolds. In general, the energies of the peak values of these curves provide estimates of the energy of the discrete state. We also use the width of these Gaussian fits as an approximation to the width of the Fano resonance, which is related to the hybridization of the individual discrete states with their respective continua, and yields the error bars in Fig.~3(b) of the main text. As can be seen when the antibonding state enters the gap, the width of the peak is dramatically reduced, corresponding to the lack of a Fano resonance. Also visible is the increasing contribution near the VBM of the bonding state Fano resonance, which in turn yields the three distinct regimes of exciton character described in the main text.

\subsection{Heuristic Excited-State Potential Energy Surfaces}
The simplicity of the Fano model allows us to easily construct approximate excited-state PESs from which we can learn combinations of relevant energy levels that lead to low activation energies. These relevant energies are, as described in the main text, the valence band maximum (VBM) and conduction band minimum (CBM) of the catalyst relative to vacuum, the Fermi energy, and the isolated orbital energies of the active site and reactant atoms. Construction of an approximate excited-state PES from these parameters requires (1) a ground state surface and (2) the decay curves of the antibonding and bonding states. One can directly obtain the ground-state PES from inexpensive DFT calculations, and even qualitatively model it with a simple Lennard-Jones potential parametrized by the desorption energy $E_d$ and the equilibrium distance $z_0$.

To obtain the excited-state PES, we need to model the $z$-dependence of the \emph{absolute} quasiparticle energies of the bonding and anti-bonding states. We expect them to vary exponentially with $z$ given the exponential localization of the relevant atomic orbitals in a diatomic molecule, as understood from Fig.~3(b) of the main text. These exponentially decaying quasiparticle levels can be qualitatively captured with a simple functional form, $\sim E_0 \pm A e^{-\kappa (z - \Delta z)}$, involving, in the simplest case, the same few parameters for the bonding ($-$) and antibonding ($+$) curves. Their asymptotic values are set by the ionization energies $E_0$ of the active site and adsorbate. The amplitude $A$, spatial decay constant $\kappa$, and shift from the origin $\Delta z$ can also be determined from a DFT calculation -- or left as free parameters  when searching for optimal catalysts.

\begin{figure}[H]
    \centering
    \includegraphics[width=.65\textwidth, trim={0cm 0 0cm 0},clip]{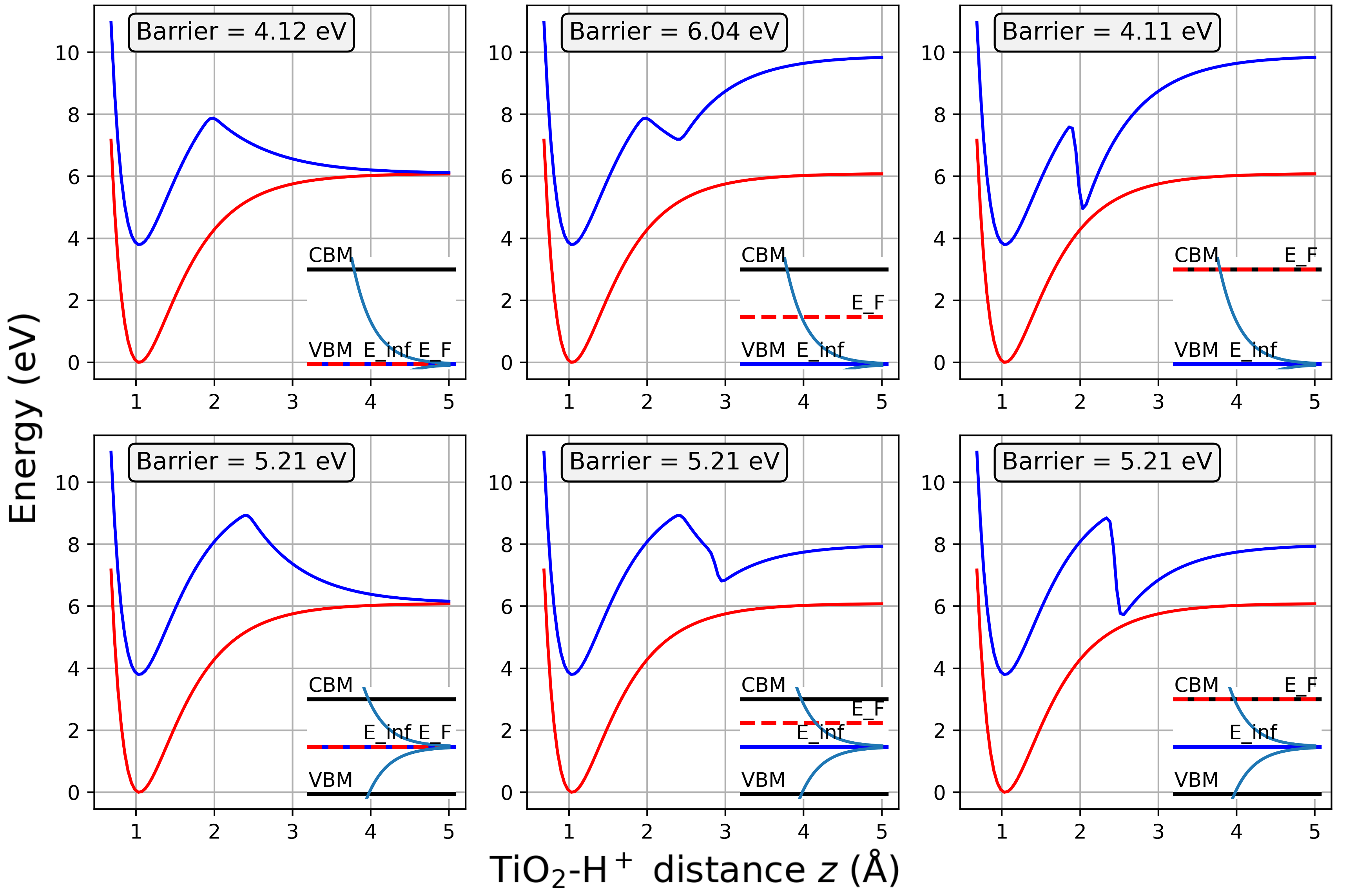}
    \caption{Examples of possible excited-state potential energy surfaces resulting from the Fano model with different parameters. Insets in the lower right corner of each plot show the relative alignment of the free parameters --- \helvet{E\_F} represents the Fermi level and \helvet{E\_inf} represents the isolated orbital level of the adsorbate and active site ($E_0$ in the above text), which for H and O is approximately equal. The evolution of the bonding/antibonding orbitals is also shown as the exponential decay curves, fitted to the data presented in Fig~3(b) of the main text. The activation barrier in the excited state is printed in each plot.}
    \label{fig:example_xpes}
\end{figure}

Once these three curves -- ground-state and bonding plus antibonding energies -- are determined, the excited-state PES is constructed by calculating the excitation energy from the alignment of the VBM, CBM, and bonding/antibonding curves, and then adding to the ground-state surface. In particular, for each $z$, one needs to determine the quasiparticle excitation edge by looking at the highest-energy occupied state between the VBM and the bonding state, and the lowest-energy unoccupied state between the CBM and the antibonding state. In Fig~\ref{fig:example_xpes}, we present a few examples of different types of PESs that can be obtained with different relative alignments of the aforementioned energies. We emphasize both the diversity of qualitative shapes and activation energies, and the degeneracy of activation energies possible with very different PESs.

The power of this approach is three-fold. First, if DFT calculations are available, one can rapidly assess whether a particular catalyst-adsorbate pair is likely to have favorable photocatalytic properties. Second, the parameters introduced above can be left free, and data-driven approaches such as data filtration or machine learning can be used to find combinations of parameters that yield low activation energies in the excited state. A particularly straightforward way to do this is with a clustering algorithm such as k-means, once a large dataset of excited-state surfaces is generated. Simple supervised clustering can be performed by applying boolean masks to the dataset conditioned on a threshold value for the activation energy and then applying the clustering algorithm. The resulting clusters represent regions in parameter space that yield low activation energies. When we performed this analysis we found the clusters to reproduce the design principles already found from our \textit{ab initio} calculations as described in the main text. Finally, the model can be applied to more complicated reactions simply by considering the relevant energy levels during each elementary step. More atoms involved would introduce more energy levels, but any bond-breaking-like event can be thought of in similar terms to the desorption reaction studied here. We leave further analysis to future works. 

\section{Crystal Structures}
Below are crystal structures for calculations presented in this work in the extended XYZ format. The ``$z$=" entry in the ``Properties" header of each structure gives the reaction coordinate $z$. All values are given in \AA.
\lstinputlisting{all_tio2_structs.extxyz}

\printbibliography[heading=bibintoc, title={References}]
\end{refsection}

\end{document}